# Swelling Mechanism of Lattice with the Ingrowth of the Defects in $UO_2$

Seçkin D. Günay[*]

Yıldız Technical University, Department of Physics, Faculty of Science, Esenler, 34210, Istanbul, Turkey

Swelling of uranium dioxide with ingrowth of defects by irradiation is not fully understood. Experimental and theoretical groups have attempted to explain this phenomenon with various complex theories. In this study, experimental lattice expansion and super saturation of the lattice were well reproduced by molecular dynamics simulation method. From the resemblance with the experimental data, here it is manifested that only oxygen frenkel pairs were created in the fission induced lattice while alpha particle radiation causes both oxygen and considerable amount of uranium defects. Moreover, in this work, defects are divided into two sub-groups as obstruction and distortion and it is shown that obstruction type frenkel pairs merely responsible for the lattice swelling for both fission and alpha particle radiation. Evidently relative lattice expansion varies linearly with the obstruction type of survived uranium defects. Additionally, at high concentrations, some of the obstruction type uranium frenkel pairs forming double or triple structures with oxygens in their octahedral cages which increase the slope of the linear dependence.

## 1. Introduction

---

[*] Corresponding author. E-mail: sgunay@yildiz.edu.tr

Uranium dioxide ($UO_2$) is, mainly used as nuclear fuel, usually involved in extreme conditions. For instance, it is exposed to operative temperatures up to 2000°C within corrosive and radioactive surroundings. $UO_2$ has been widely studied in order to understand thermo physical, transport and defect properties (1-14). All these are comprehensively summarized by Govers et. al. 2007 (15,16). Molecular dynamics simulation of the temperature effect on physical properties of $UO_2$ has assisted the ongoing experiments which cannot be done easily because of extreme conditions. Not only the temperature but also irradiation has the effect on the lattice parameter, volume, density, electrical resistivity and diffusion are considered among these properties (17-22). Radiation damage in $UO_2$ crystal has great influence on reactor fuel that causes the degradation of performance.

Heavy ions, fission products, alpha particles, alpha-recoil atoms (alpha particle and recoil nucleus produced in alpha decay) and neutrons are the reasons for irradiation damage. Moreover electrons, X-rays and gamma rays would enhance damage but generally ignored (17). When an $UO_2$ crystal exposed to radiation, frenkel pair (FP) defects are created in the direction of the radiation path. FPs are lattice vacancies and the atoms which jumped from the lattice site to an interstitial position by the enforcement of irradiation. Furthermore, if the implemented dose is increased, complete amorphization of the crystalline may occur at ambient condition (17). Experimentally, Nakae et al. (18-20) and Weber (21,22) have studied the fission and α-particle dose effect and temperature dependence of lattice parameter, lattice strain and their recovery behavior of irradiated $UO_2$.

There have been radiation damage molecular dynamics simulation studies in which defect production and clustering by energetic uranium recoils in $UO_2$ have been investigated (23). In addition, Aidhy et.al. (24) have explored the kinetic evolution of irradiation-induced point defects in $UO_2$ by molecular dynamics simulation at 1000K. They have observed that if the defects are present in only one sublattice, the FPs recombine during equilibration, whereas if defects are present in both sublattice they form clusters and conclude that the cation sublattice is primarily responsible for the radiation tolerance or intolerance of material. However, to our knowledge, computer simulations of the lattice swelling with the defect ingrowth and the lattice recovery with temperature in defected $UO_2$ have not been considered so far.

In the present study, molecular dynamics simulation calculations were carried out for the supercell of $UO_2$. The defected samples were prepared according to the number of defects which correspond to the suggested experimental dose. Two different type of partially ionized rigid ion potentials, existed in literature (1,5), were used for the interionic interactions. The ingrowth of defect versus swelling of the lattice is investigated and the results are compared with the experimental data.

**2. Molecular Dynamics Simulation**

2.1. *Procedure*

Crystalline uranium dioxide with four uranium and eight oxygen ions in its unit cell has the fluoride type structure. Every uranium ion is at the center of surrounding cube and coordinated to eight oxygen anions. Oxygen is surrounded by four Uranium ions. MD cell is constructed by setting 500 cations and 1000 anions with array 5x5x5 supercells in five mutually orthogonal directions. The calculations have been carried out by the MD code Moldy (25). Long-range coulomb interactions are accounted with Ewald's summation (26). The positions and velocities of the ions are calculated by Beeman's algorithm which is predictor-corrector type, with the time step $\Delta t=1.0$ fs. The system has been simulated at constant pressure and temperature (NPT) ensemble at 300K by applying the Nose-Hoover thermostat and Parinello-Rahman constant stress method. Equilibrium run was performed for 10 ps and then the data were accumulated over the following 40 ps.

*2.2. Sample Preperation*

The irradiated samples of 5x5x5 supercell with the different defect concentrations have been prepared by randomly replacing an ion from its lattice site to an interstitial position on the layers. A representative sample of defected supercell is given in Figure 1. Here after, through this article, these ions are called initial frenkel pair (IFP) defects. After the equilibration, they are called frenkel pair (FP) defects. To take into account the annihilation effect, IFP defects have been created such that one layer is defected at every two layers, so nearby vacancies and interstitials do not terminate each other directly. Each defected layer has approximately the same number of IFPs. The simulation procedure for the defected and perfect supercell boxes is the same. Samples of supercell boxes of uranium dioxide with IFPs are prepared for several different defect concentrations, based on the experimental dose (17). Supercell boxes are built up with oxygen IFPs or uranium IFPs but not both together in the same sample in order to correlate the sublattlice effect with irradiation type. The interionic interactions have been modeled by using two different types of rigid ion pair potentials parameterized by Yakub et. al.(5) and Günay et. al.(1)

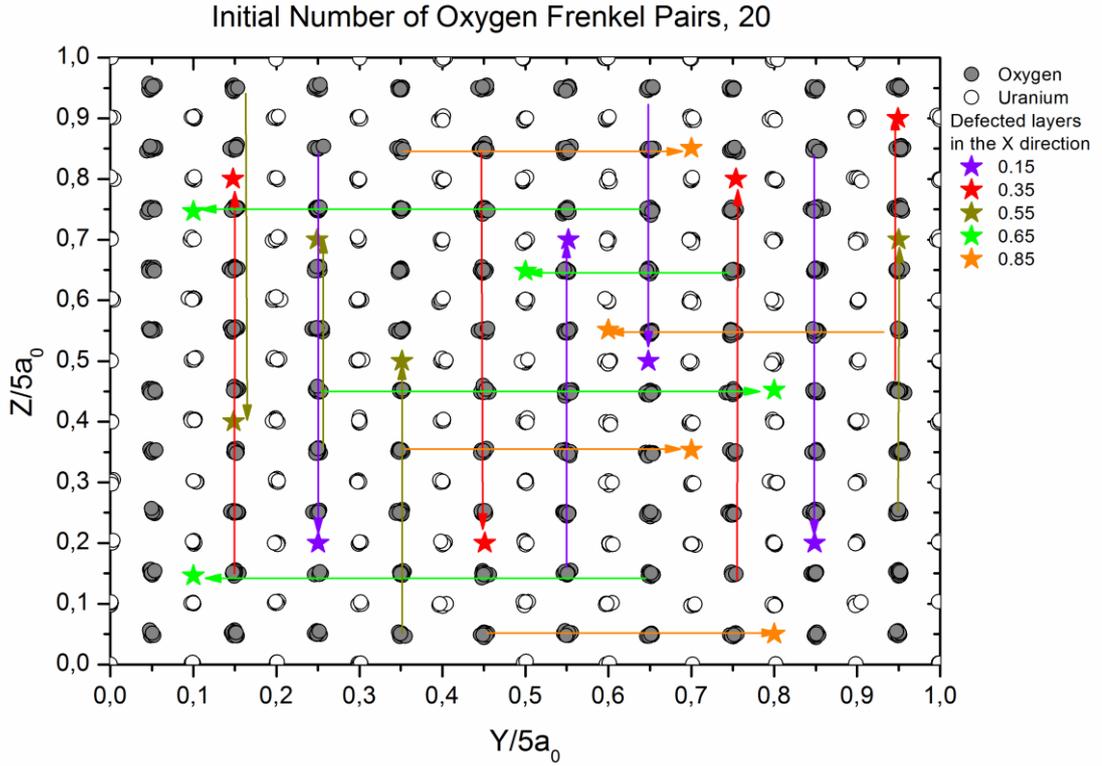

**Figure 1.** Sample view of initial defected MD supercell with 20 oxygen IFPs from x direction. Arrows indicate the displacements of atoms from their lattice sites to interstitial positions. Each color represents the different layer in x direction.

2.3. *Calculation of number of defects*

In this part, the number of both oxygen and uranium FPs in the prepared samples are calculated. Although it is possible to determine the number of defects visually from different directions using the VMD program (27) as it can be seen from the Figure 2 we have developed a method to calculate the average number of FPs. The results obtained from the method are consistent with the visual observations.

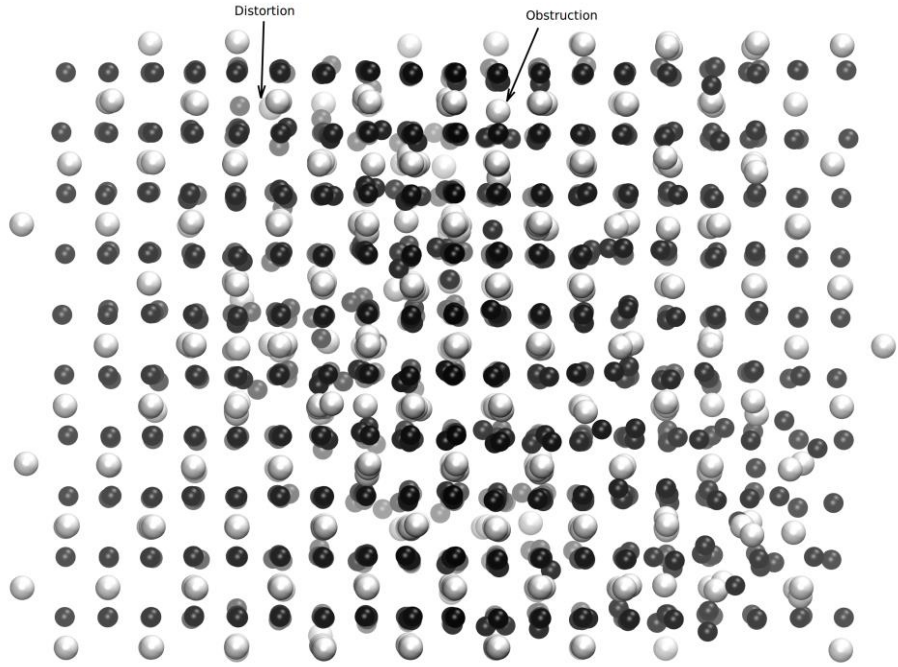

**Figure 2.** The equilibrium view of $UO_2$ supercell initialized with 17 uranium IFPs from <110> direction. The distortion and obstruction type uranium (gray) defects are indicated with arrows. The distortion type is considered as small displacement of an ion into the channel. The obstruction type is considered as occupation of the central positions into the channel by an ion; it has a special position with a constant coordination number that can be considered as an ion trapped in a cage. There exist a relation with the ingrowth of obstruction type defects and the lattice expansion with irradiation.

In this method, the number of defects are calculated by averaging over the number of time steps, $n_t$, with the following relation,

$$\overline{M_\alpha} = \frac{\sum_{i=1}^{n_t} \sum_{j=1}^{N_\alpha} K_{ij}}{n_{\alpha\beta} n_t} \tag{1}$$

where $K_{ij}$ represents the number of $\beta$ ions around the $j^{th}$ $\alpha$ ion at the $i^{th}$ time step within the range $\Delta r = r_{max} - r_{min}$ and $n_{\alpha\beta}$ is the first coordination number of perfect crystal $n_{uu} = 12$, $n_{oo} = 6$, $n_{uo} = 8$. $r_{min}$ and $r_{max}$ values are determined from the first peak of the radial distribution function $g_{\alpha\beta}(r)$ with the constrains that $\overline{M_u} \cong N_u = 500$ and $\overline{M_o} \cong N_o = 1000$. As a representative sample, $g_{uu}(r)$ is given in Figure 3 that is used to determine the maximum and the minimum values of r. The distortion and obstruction type defects (see Figure 2) can be, respectively, considered as displacements of the ions into the channels and occupation of the central positions of the channel by the ions. Obstruction type ions have a constant number of ions surrounding it but distortion type ions are much more mobile. It is observed in Figure 3 that the distinct pre-peak just before the principal peak reflects the distribution of the obstruction type of FPs.

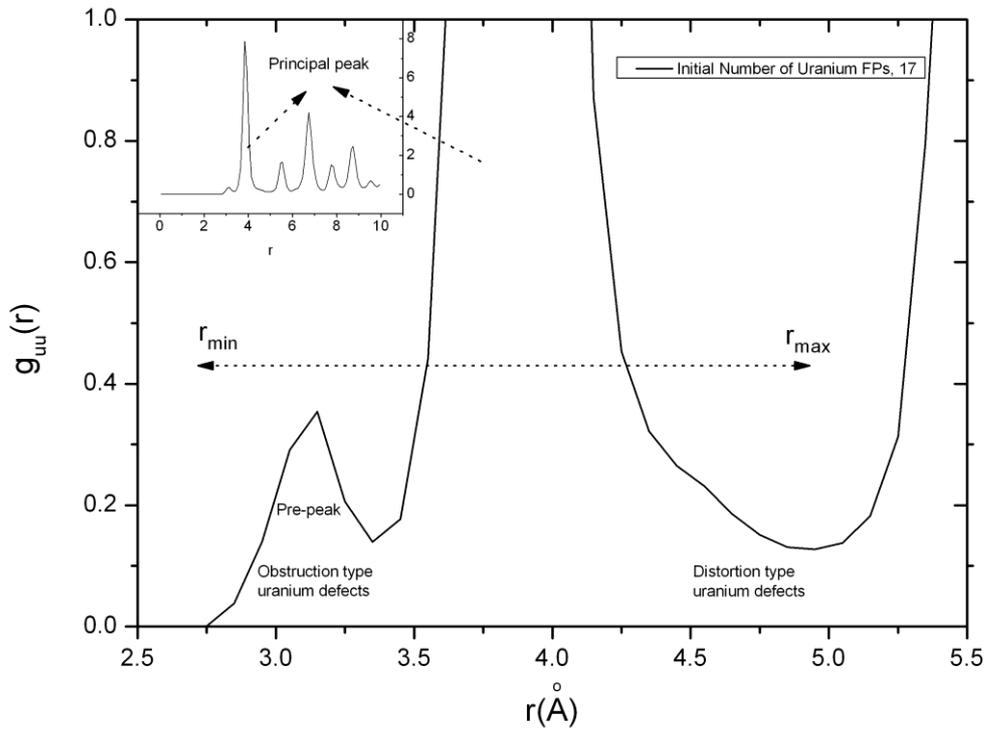

**Figure 3.** Radial distribution function $g_{uu}(r)$ of $UO_2$ with 17 uranium IFPs for the 5x5x5 MD supercell. Obstruction and distortion type defects are indicated on the graph. Obstruction type defects have sharper boundary (pre-peak) than the distortion type.

The Eq. (1) gives the average number of obstruction type of defects $\overline{M_u}^{obs.} \cong 14$ for the range $\Delta r \cong 3.4 - 0.0$. To calculate the distortion type of FPs, $\Delta r$ is determined by the assumption that the principal peak of $g_{\alpha\beta}(r)$ is symmetrical. It is a little ambiguous case to determine an interval for distortion type defects than obstruction type defects. Here again radial distribution function is helpful. When it is determined, $\overline{M_u}^{dist.}$ is calculated as about 21 for $\Delta r \cong 4.947 - 4.5$. These calculated numbers of defects are consistent with the visual observation of the VMD snapshot given in Figure 2.

**3. Results and Discussions**

3.1. *Swelling of the lattice with ingrowth of defects*

The difference between the lattice expansion with fission damage and alpha particle damage is about a few orders of magnitude (17) where the fission damage is less effective than the alpha particles. The relative lattice expansion $\Delta a/a_o$ has been calculated at 300K for the IFP defect numbers varying up to 40 for 5x5x5 supercell system. Comparing the results with experimental relative lattice expansion have shown that the calculations carried out with the defected sample separately with Oxygen IFPs and Uranium IFPs resembles the experimental results with fission and alpha particles, respectively. Therefore, it is more convenient to present the results in two separate sections.

3.2. *Defected Supercell with Oxygen Frenkel Pairs*

Fig 4. presents the number of IFPs versus relative lattice expansion for Günay(1) and Yakub(5) potentials together with the experimental relative lattice expansion with fission dose taken from Matzke (17). MD simulation for NPT ensemble has been run for 10 different initially defected systems with the number of oxygen IFPs varying up to 40. During 10 ps equilibration process only some of the oxygen IFPs were able to survive due to recombination of interstitials and vacancies. The number of survived FPs, which are indicated with the numbers near the data points in Figure 4, remains constant during the simulation. As the fission dose increased, it has been observed that three separate stages occur experimentally (17, 18): production of isolated FPs at constant rate $(14<\log(F)<16)$, newly produced displaced atoms are trapped at existing defects resulting with interstitial clusters $(16<\log(F)<17.5)$ and no more new interstitials are produced that is called the super saturation stage $(17.5<\log(F)<18.5)$.

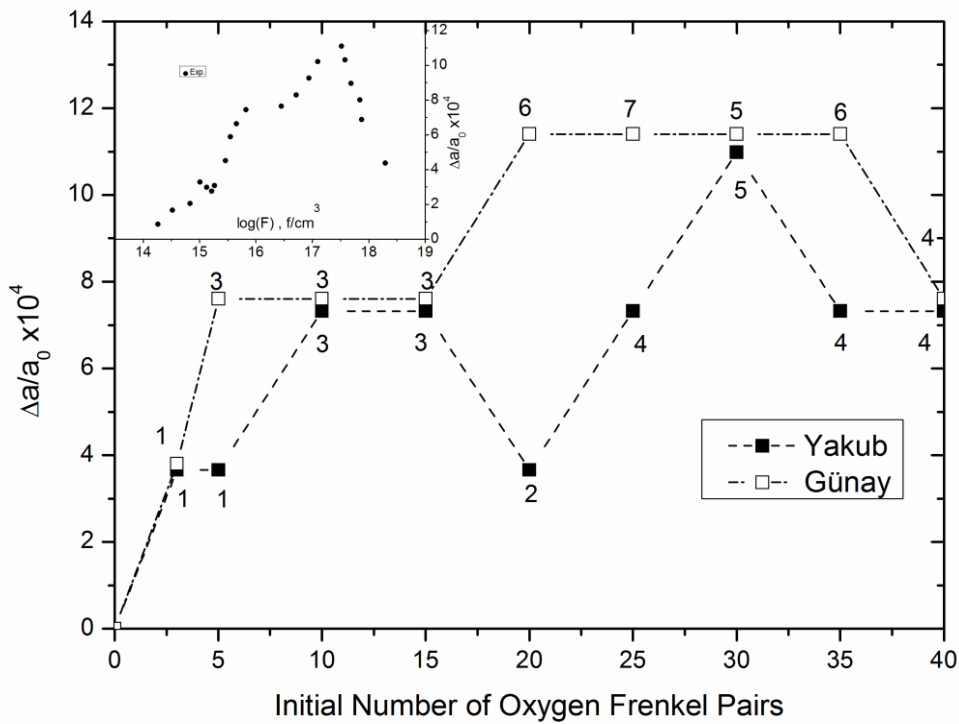

**Figure 4.** Lattice change versus initial number of oxygen frenkel pairs in the MD supercell. The values given near the data points represent the number of survived obstruction oxygen FPs in the MD supercell. The inset is taken from Ref. 17 and shows the experimental data of change of lattice parameter as a result of ingrowth of defects with increasing fission dose

These stages are successfully reproduced for both potentials. The maximum swelling point is estimated at $\Delta a/a_o = 11.4 \times 10^{-4}$ which is very close to experimental value. From the simulation data, it could be commented that there might exist another stage which is between $14 < \log(F) < 15$. Consequently four of the stages are almost equidistant in y direction which might be related with the interstitial oxygen diameter. Also the relative lattice expansion is proportional with the number of obstruction type of the survived oxygen FPs. The volume increment per obstruction type of oxygen FP $\Delta v_F$ is estimated from the following relation (17)

$$\frac{\Delta a}{a_o} = \left(\frac{\Delta v_F}{3n_{cell}V_o}\right)N_F \qquad (2)$$

where $n_{cell}$ is the number of unit cells in the supercell, $V_o$ is the unit cell volume and $N_F$ is the number of FPs as approximately about $10 \overset{o}{A}^3$ and $13.43 \overset{o}{A}^3$ for Günay and Yakub potentials respectively. Nakea et al. (18) have estimated the volume increments associated with an interstitial is about $13.8 \overset{o}{A}^3$ without indicating any specific sublattice (uranium or oxygen). But in this work, it can be deduced that, the change in the number of oxygen FPs (e.g. by fission irradiation) does not affect the uranium sublattice which remains almost perfect without any defect. According to this observation, it can be interpreted that the experimental fission dose can only create oxygen FPs which are responsible for the lattice swelling.

## 3.3. *Defected Supercell with Uranium Frenkel Pairs*

The aim of this part is to correlate relative change of lattice parameter of alpha irradiated $UO_2$ with MD simulation of defected $UO_2$ supercell with uranium FPs and to interpret the results. This section is separated into two subsections because fraction of defects is an order of magnitude higher than oxygen IFPs part. MD simulation cell is built for both 5x5x5 and 8x8x8 unit cells in order to observe the influence of defects images which is caused by periodic boundary conditions. Results for 5x5x5 supercell and 8x8x8 supercell with Uranium IFPs are compared so that they could be confirmed.

### 3.3.1 5x5x5 Supercell

The number of uranium IFPs has been varied up to 30 which were enough to observe the saturation stage for both Yakub and Günay potentials. Contrary to the defected $UO_2$ with oxygen IFPs, during the equilibration process, recombination of some of the uranium vacancies and interstitials has been resulted with the creation of the defects of distortion and obstruction type of oxygen FPs and obstruction type of uranium FPs. Figure 5 and 6 show the variation of the number of survived (uranium) and created (uranium and oxygen) FPs with uranium IFPs counted from <110> direction as in Figure 2. All types of defects increases to a saturation value. The main contribution to the total number of defects comes from the distortion type of oxygen defects which shows sharp increase to the saturation value. Similar behavior was observed by Turos et. al.(28) in their experimental study of radiation defects in $UO_2$. For comparison, the inset in Figure 5 and Figure 6 show the variation of the U and O defect concentrations with increasing implantation dose. It is interpreted that the large difference in the steering force of U and O rows causes the higher value of the saturation value for oxygen ions than that of uranium ions. This force arises from the mass and mobility difference between U and O ions. It is known that O ions are, naturally, more mobile than U ions and have liquid like self diffusion coefficient at about 20% below the melting temperature that makes the $UO_2$ to be an oxygen superionic conductor. Therefore, oxygen defects have been easily created during the recombination of U IFPs even at 300K.

The experimental channeling yield ratio has been given as $\chi^O/\chi^U \cong 6.7$ (28). Analogously, we estimate the ratio of the total defect numbers (distortion and obstruction type) of oxygen ions to that of uranium ions as about 3.3 for Yakub potential and 5.4 for Günay potential. Our results show that because of the larger effective ionic radii of oxygen ions $r_O/r_U \cong 1.4$, distortion type defect of an oxygen ion also obstructs the channel, even that it is not in the centre, and give rise to an enhanced scattering yield that observed experimentally (28).

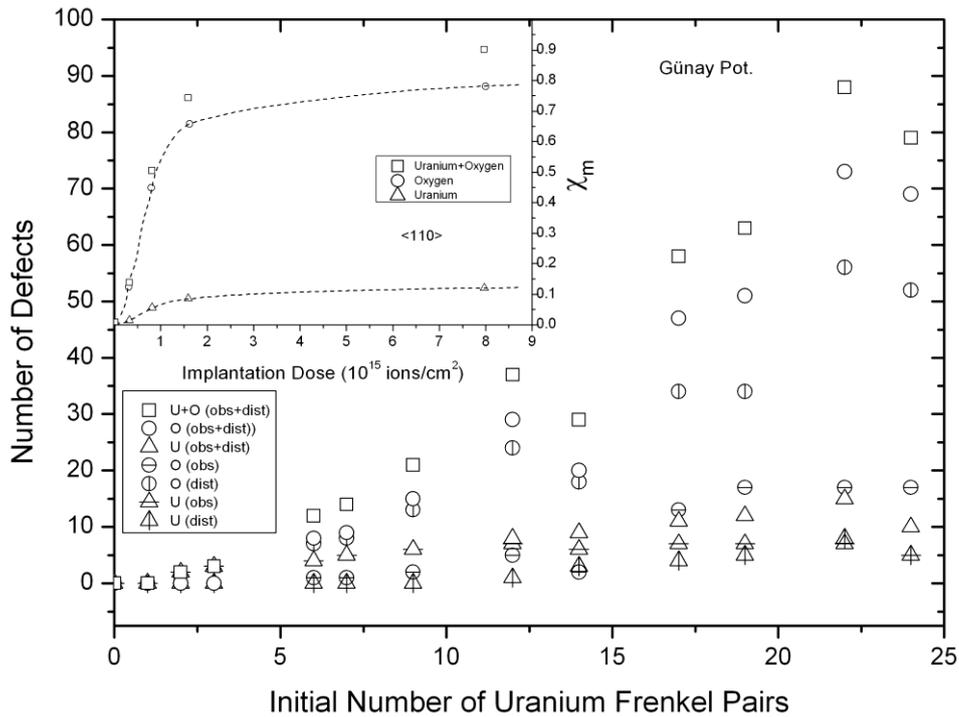

**Figure 5.** Variation of the number of survived and created FPs with the number of uranium IFPs using Günay potential. The inset is taken from Ref. 28 and show the experimentally estimated concentrations of uranium and oxygen FPs created by the dose.

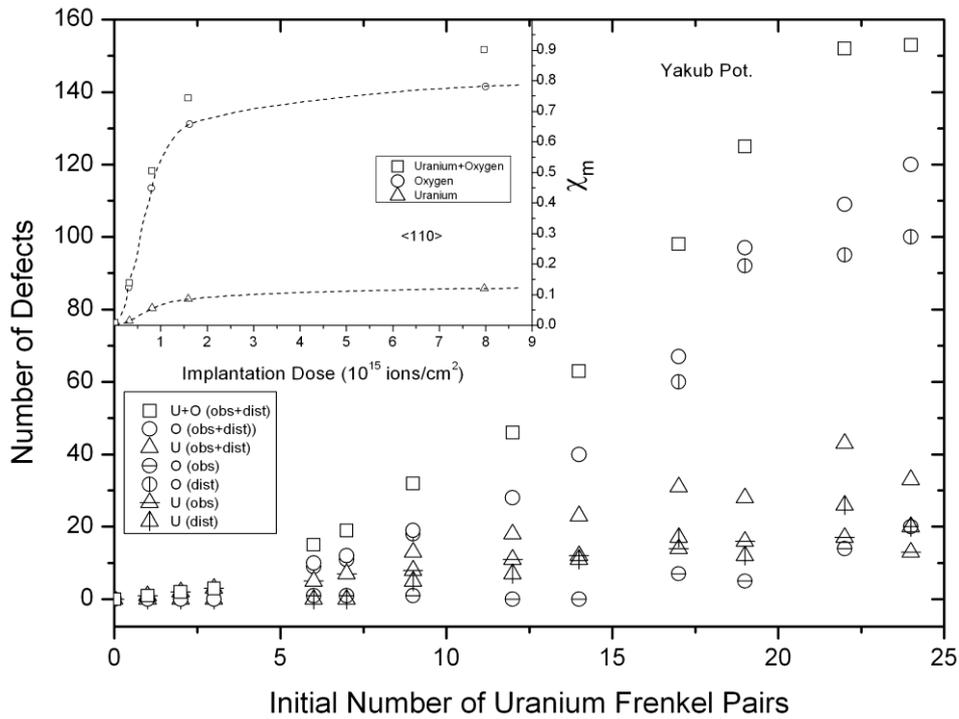

**Figure 6.** Variation of the number of survived and created FPs with the number of uranium IFPs using Yakub potential. The inset is taken from Ref. 28 and show the experimentally estimated concentrations of uranium and oxygen FPs created by the dose.

Additionally, distinct feature in Figure 5 and Figure 6 is that further increase in the uranium IFPs causes the reduction of the number of obstructed and distorted uranium and oxygen ions. Such tendency of the recovery of the defects was not the case in the experimental studies. The question is that would the further increase of dose in experiment have been resulted with the partial recovery of the defects and contraction of lattice?

Figure 7 and 8 present the lattice expansion $\Delta a / a_0$ with the number of defects. There is no, clear, functional dependence of $\Delta a / a_0$ with the obstruction type of oxygen defects and the distortion type of uranium defects which have, respectively, almost no contributions to the total defect up to about ~15 and ~10 uranium IFPs. Therefore, these are not included in Figure

7. Until the saturation level, the lattice expansion has an exponential dependence to the number of distortion type of oxygen (see Figure 7(a)), distortion+obstruction types of oxygen (see Figure 7(c)), distortion+obstruction types of oxygen+uranium (see Figure 8(a)) and initial uranium FP (see Figure 8(b)). The maximum lattice expansion observed is about 1.4% for the Yakub potential and 0.5% for the Günay potential which are respectively correspond to volume change of 4.2% and 1.5%. For comparison, the experimental lattice expansion with alpha dose by Weber (21) is also shown as inset in Figure 8(b). Fitting of the data to the damage ingrowth model that yielded to the expression as $\Delta a/a_0 = 8.4\times 10^{-3}\left[1-\exp(-0.85 D_\alpha \times 10^{-16})\right]$, has predicted the lattice expansion as 0.84% at saturation (21). Inspiring with Weber's equation, we have also fitted the data to the same equation and in all cases the saturation value of the lattice expansion $\Delta a/a_0$ is estimated between $0.44\% - 0.48\%$ for Günay and $1.3\% - 1.8\%$ for Yakub potentials. The reason of obtaining lower lattice expansion might be the case of Günay potential is the stronger attractive interaction in their model potential compared to that of Yakub. Weber (21) has estimated that one defect pair for every 3 to 4 unit cell occurs at the saturation region and has suggested that such concentration of defects signals isolated defects and negligible clustering. Both potentials give almost the same number of obstruction type defects as Weber had found. It should be also noted that lattice expansion in Figure 7(b) varies linearly with the obstruction type of survived uranium defects which is interesting. Evidently, these uranium defects are the reason for lattice expansion which is coordinated with six uranium ions. According to Eq. 2, the tangent of Figure 7(b) gives volume increment per obstruction type uranium frenkel pair, $\Delta v_F$, for Günay and Yakub potentials, as $39.56 \overset{o}{A}{}^3$ and $49.47 \overset{o}{A}{}^3$, respectively.

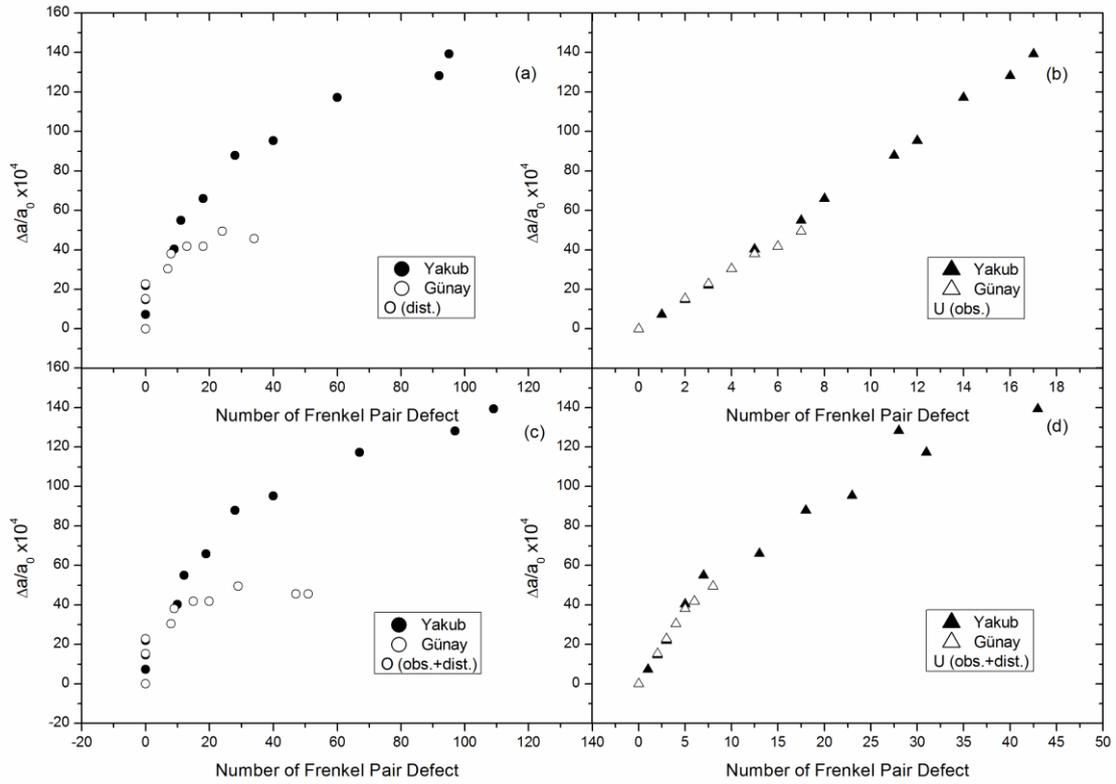

**Figure 7.** Relative lattice expansions with different types of defects are shown as circle for oxygen and triangle for uranium: (a) Distortion of oxygen (b) obstruction of uranium (c) obstruction+distortion of oxygen and (d) obstruction+distortion of uranium.

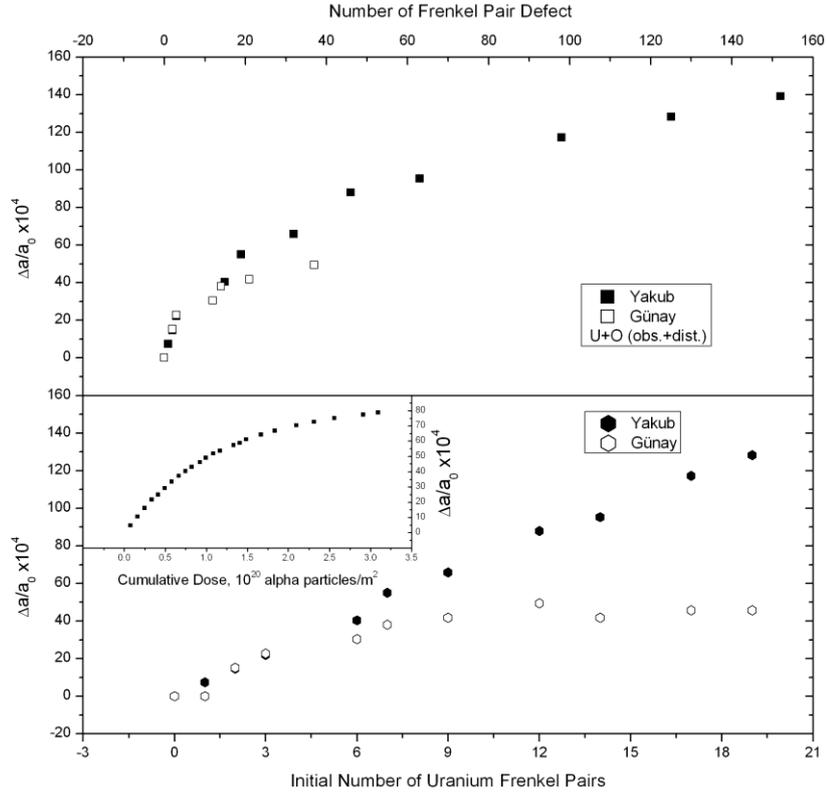

**Figure 8.** Relative lattice expansions with (a) total number of defects after equilibration and (b) initial number of uranium FPs. The inset is the experimental lattice expansion with cumulative dose taken from Ref. 21.

*3.3.2 8x8x8 Supercell*

Here similar procedure is used as the 5x5x5 supercell. Uranium IFP defects are created up to 80 for the supercell of 2048 uranium and 4096 oxygen ions. For every defect created in the supercell equilibrium run is performed for 30ps and data is accumulated for 70ps. When the ions are visually analyzed, there is a clear distinction between oxygen defects and uranium defects. Uranium defects, which are generally obstruction type, have their stable positions at octahedral interstitial sites coordinated with other 6 uranium ions which have retained their normal positions in the crystal. When the crystal is examined from <110> direction it is clear to see that obstruction type uranium defects obstructed the channel without changing their interstitial position and vibrating at their sites. In contrast to these, oxygen defects, which are generally distortion type, are mobile without any stable interstitial position. Moving from one crystal position to another crossing the channel or partially closing the channel (displacement of oxygens into the channel) without any special positions. These can be deduced either from images or from the radial distribution function of O-O and U-U.

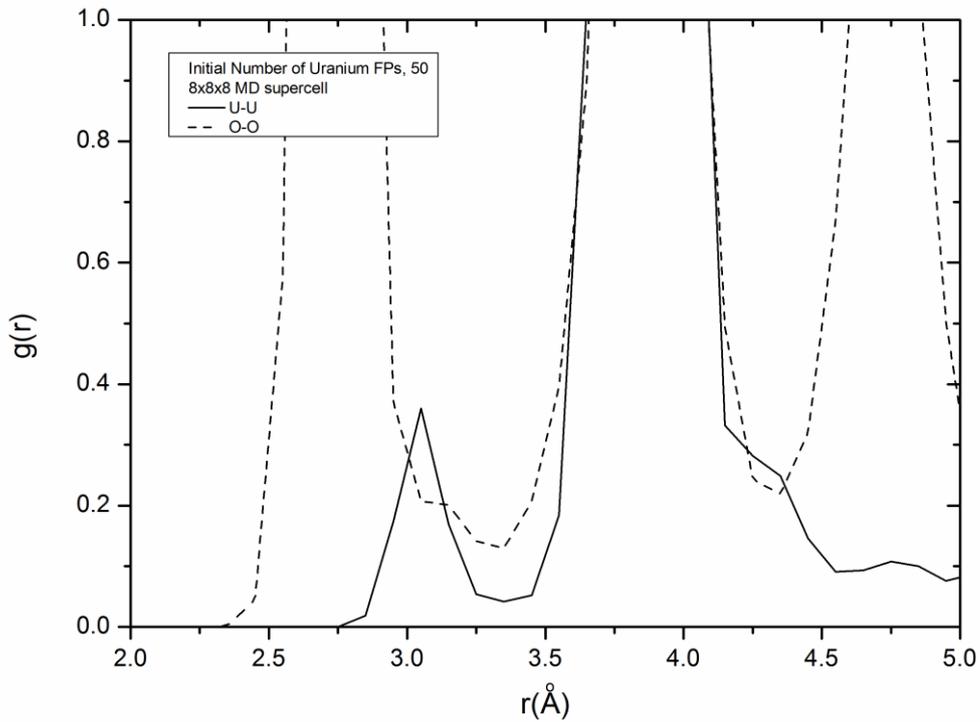

**Figure 9.** Radial distribution function $g_{uu}(r)$ and $g_{oo}(r)$ of $UO_2$ with 50 uranium IFPs for the 8x8x8 MD supercell.

Uranium defects have their interstitial positions between $2.75 - 3.4 \, \overset{o}{A}$ surrounded by uranium ions (Figure 9). Sharp peak indicates the immobile ions. Oxygen defects can be found between $3.2 - 3.4 \, \overset{o}{A}$ without any significant peak. Only very small amount of oxygens are immobile, they could be observed from <110> direction and retain their positions in the channel which is called obstruction type oxygen defects. Associated peak of these could not be observed from the Figure 9 which might have been merged with the distortion type. Obstruction type oxygen defects are relatively much less than distortion type oxygen defects and can only be observed above 40 IFP uranium ions.

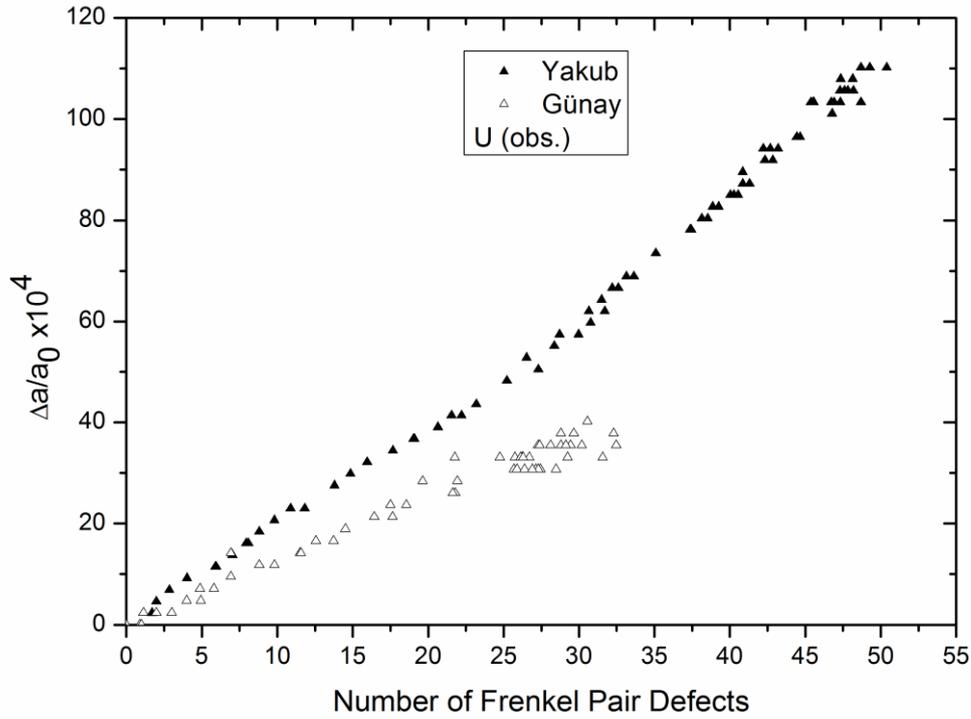

**Figure 10.** Relative lattice expansions versus number of obstruction type of uranium defects for Yakub and Günay potential.

In Figure 10, linear dependence of obstruction type uranium defect number to swelling of the lattice clearly observed for both potentials. This interesting feature explains the reason of lattice expansion during the alpha particle irradiation. This is also observed in previous section for 5x5x5 supercell so that the results are confirmed. Volume increment for obstruction type uranium FPs from Figure 10 is approximately about $55.65 \overset{o}{A}{}^{3}$ for Yakub and $28.85 \overset{o}{A}{}^{3}$ for Günay potentials. It is an approximate value because in Figure 10 the slope of the graph is slightly different above and below 30 obstruction uranium FPs for Yakub potential. A possible reason for this is, obstruction type oxygen defects appear above 35-40 uranium IFPs which corresponds to 30 obstruction uranium FPs in Figure 11.

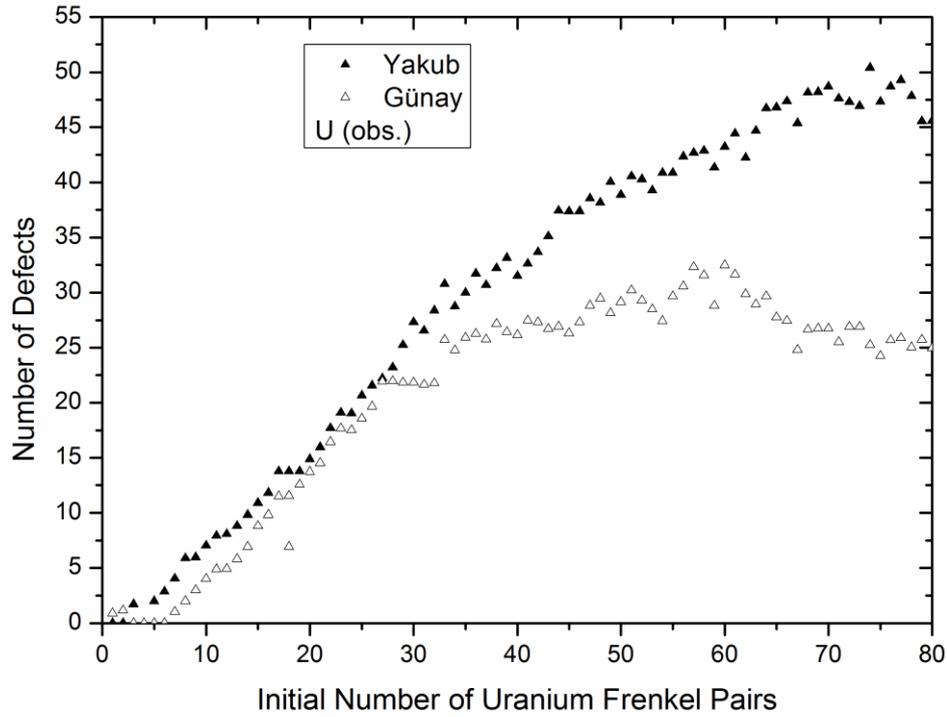

**Figure 11.** Number of uranium FPs versus the number of uranium IFPs for Yakub and Günay potential.

In the previous chapter, it is also observed in Figure 6, obstruction type oxygen FPs appear above 15 uranium IFPs. As we mentioned in the previous section obstruction type oxygen defects are only a very small fraction of oxygen defects. From the slope of Yakub potential graph in Figure 10, for the interval 0-30 uranium FPs, volume increment is calculated $46.62 \overset{o}{A}^3$ and for the interval 30-50 uranium FPs, volume increment is calculated $64.97 \overset{o}{A}^3$. Difference between these gives $18.35 \overset{o}{A}^3$ which is close to the value of volume increment for oxygen FPs that is calculated in section 3.2 as $13.43 \overset{o}{A}^3$ for Yakub potential and $13.8 \overset{o}{A}{}^{3}$ for the experimental result (18). These outcomes give the idea that small amount of oxygen defects penetrate into the octahedral cages of obstruction type uranium defects. It could be observed that some octahedral cages are hosting one or sometimes two oxygen atoms and these oxygens are vibrating at this interstitial site as an obstruction type defect. Not all of the obstruction type oxygens involve in this incident. This event, which can also be observed visually, result an increase in the slope in Figure 10 above 30 uranium FPs.

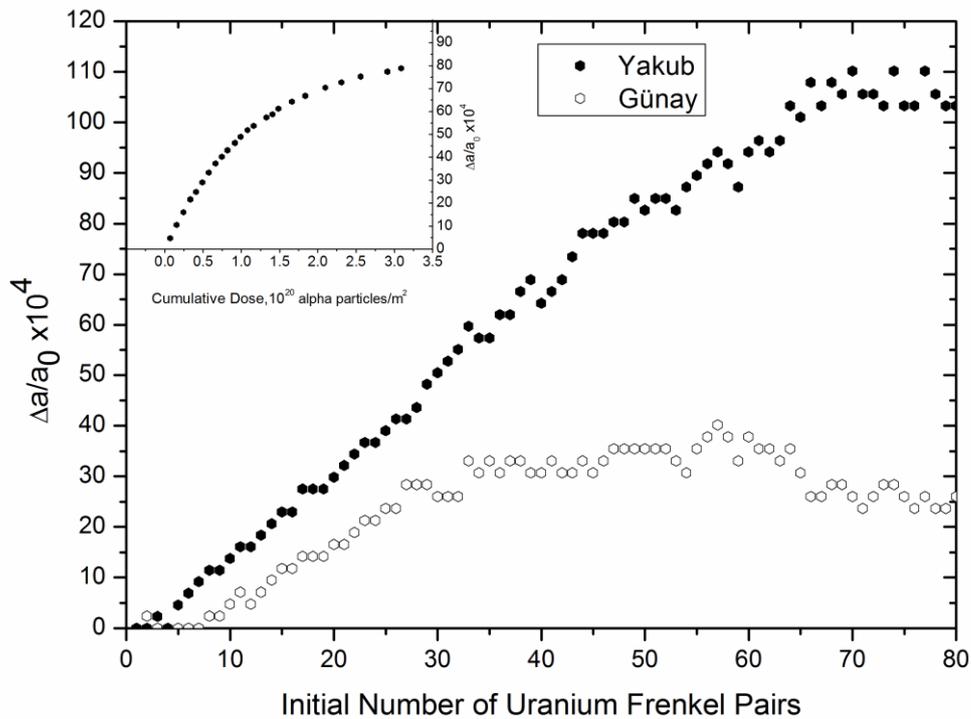

**Figure 12.** Relative lattice expansions with the number of uranium IFPs for Yakub and Günay potentials. The inset is the experimental lattice expansion with cumulative dose taken from Ref. 21.

In Figure 12, when obstruction type oxygens appear above 40 up to 80 uranium IFPs, swelling of the lattice is exponentially decaying with an increasing form, saturated to an upper value. Here it might be speculated that oxygen and uranium ions, which existed in the octahedral cage as a couple, at some point collapse results a decrease or recombine results an increase in the lattice so that the high scattering of the data above 40 in Figure 12 is observed. When obstruction type uranium ions are created and obstruction type oxygens combine more and more with obstruction type uranium ions, swelling of the lattice reaches a saturation point until no space left for the obstruction defects. These saturation points are $0.3-0.4\%$ for Günay and $1.0-1.1\%$ for Yakub potential. As the supercell size is increased from 5x5x5 to 8x8x8, saturation value of Yakub potential gives better agreement with experimental value $0.84\%$.

## 4. Conclusion

Molecular dynamics simulations have been performed to investigate swelling of the lattice with the ingrowth of defects in $UO_2$ using two different type of partially ionized rigid ion potentials. Defected supercells for different defect concentrations have been simulated in constant pressure-temperature ensemble. Some of the IFPs, for both oxygen and uranium defected samples, have been recombined during the equilibration process that induces additional obstruction and distortion type of FPs. The numbers of induced defects remain almost constant after equilibration.

Resemblance between lattice expansion with oxygen defect ingrowth and that with fission dose exhibits that the experimental fission damage can only create oxygen FPs which are responsible for the lattice swelling. Experimental stages and saturation values were successfully reproduced.

For the uranium IFP sample, the simulations results are similar to the experimental data obtained in the study of defect production and annealing with the high energy $^4$He ions (28). Because of the large vibrations and the size of oxygen ions, distortion type of defects may also give rise to an enhanced scattering yield even that they are not exactly occupying the center of the channel. Lattice expansion clearly varies linearly with the obstruction type of survived uranium defects while the rest have the exponential dependence. Six coordinated (U-U) uranium interstitials are the reason for lattice expansion. Although there are considerable differences between the saturation value of lattice expansions from experiment and our simulation, the estimated obstruction type of defect concentration is almost the same as experimental value. Saturation value of lattice expansion for both fission and alpha particle irradiation exist because interstitial ion volume grow to an upper limit, but at some point volume is too big for the obstruction type ion to have a stable point.

Next step of this study will concern with evolution of the lattice recovery with the temperature for the supercell with oxygen and uranium FPs. This will enable us to understand the effect of the sublattice on the recovery procedure.

**Acknowledgements**

The author gratefully acknowledges Prof. Dr. Çetin Taşseven for his helpful discussions and detailed comments to the manuscript.